\documentclass[11pt]{article}
\usepackage{CJKutf8}
\usepackage{hyperref}
\usepackage{amsmath,amssymb,color,graphics,epsfig,tikz}

\DeclareMathOperator{\sech}{sech}
\usepackage{amsfonts}

\newcommand{\be}{\begin{equation}}
\newcommand{\ee}{\end{equation}}
\newcommand{\bea}{\setlength\arraycolsep{2pt} \begin{eqnarray}}
\newcommand{\eea}{\end{eqnarray}}
\newcommand{\nn}{\nonumber}

\def\fft#1#2{{\frac{#1}{#2}}}

\def\0{{\sst{(0)}}}
\def\1{{\sst{(1)}}}
\def\2{{\sst{(2)}}}
\def\3{{\sst{(3)}}}
\def\4{{\sst{(4)}}}
\def\5{{\sst{(5)}}}
\def\6{{\sst{(6)}}}
\def\7{{\sst{(7)}}}
\def\8{{\sst{(8)}}}
\def\sst#1{{\scriptscriptstyle #1}}

\begin{document}
\begin{CJK}{UTF8}{gbsn}

\begin{flushright}
\end{flushright}

\vspace{25pt}
\begin{center}
{\large {\bf Exactly solvable complex $\mathcal{PT}$ symmetry potential $A[\sech(\lambda x)+i \tanh(\lambda x)]$}}

\vspace{10pt}
Wei Yang

\vspace{10pt}

{\it College of Science, Guilin University of Technology, Guilin, Guangxi 541004, China}

\vspace{40pt}

\underline{ABSTRACT}
\end{center}

We obtained the exactly solutions of the $\mathcal{PT}$ symmetric potential $V(x)=A[\sech(\lambda x)+i \tanh(\lambda x)]$, and found this system has no bound-state.
which $\mathcal{PT}$ symmetric potential was first studied in this article, and the handedness effect is showed from reflection coefficients. As the asymptotically non-vanishing imaginary potential component, when the direction of the incident wave is opposite, that the transmission coefficient will emerge a complex phase factor. 

\vfill {\footnotesize Emails: weiyang@glut.edu.cn}
\thispagestyle{empty}

\pagebreak



\newpage

\section{Introduction}
Since the $\mathcal{PT}$ symmetric quantum mechanics was introduced by Bender and Boettcher \cite{Bender:1998ke}, the non-Hermitian quantum mechanical has developed rapidly.
They showed that $\mathcal{PT}$ symmetric quantum mechanics can have real eigenvalues. 
In other words, we can be constructed non-Hermitian of the Hamiltonian to describe the real physical world\cite{Bender:2002vv,Bender:2007nj}. There are some topological properties in Non-Hermitian Physics can be found in several recent paper \cite{Okuma:2022bnb,Kawabata,Chen:2022lim,Ding:2022juv}.

As the $\mathcal{PT}$ symmetry parity which due to one-dimensional  Schr\"{o}dinger equation  with potential must have relation $V(x)=V(-x)^*$. 
Later $\mathcal{PT}$ symmetric quantum mechanics further extends to pseudo-hermiticity quantum mechanics \cite{Mostafazadeh:2001jk,Mostafazadeh:2001nr,Mostafazadeh:2002id}. 
The spontaneous breakdown of $\mathcal{PT}$ symmetry was discover from the potentials $V(x)=x^2(ix)^\epsilon$, studies show that when the exponent $\epsilon\geq 0$ energy eigenvalues are real and positive, but when $\epsilon<0$, the eigenvalues  will emerge complex conjugate pairs \cite{Bender:1998ke}. In addition to numerical research, there are a lot of analytical works to solve the $\mathcal{PT}$ symmetric potentials \cite{Znojil:1999qt,Znojil:2000ia,Znojil:2002yr,Ahmed:2001gz,Levai:2017vdu,Soliemani:2021lpt}. In fact,  we can construct $\mathcal{PT}$ symmetric potentials from Natanzon class \cite{Bose,Natanzon} in generally. 

Here we study the new $\mathcal{PT}$ symmetric potential $V(x)=A(\sech(\lambda x)+i \tanh(\lambda x))$, defined on the real $x$-axis, and investigate its bound and scattering states.  
This $\mathcal{PT}$ symmetric potential has not been considered in previous articles, and its imaginary component asymptotically non-vanishing. This similar potential is also considered in the article\cite{gle,lg}.
In particular, the discussion of the finite $\mathcal{PT}$ symmetric square well potential in \cite{lgj}, which is similar in scattering properties to the potential  $V(x)=A[\sech(\lambda x)+i \tanh(\lambda x)]$ and all have handedness effect, but the bound state are different, that the potential considered in this paper have no  bound state.

The work is organized as follows. 
In Sec.~\ref{Sec1}, we give the exact solutions for this potential. In Sec.~\ref{Sec2}, use the exact solution to building up scattering information.  In Sec.~\ref{Sec3},
we found there is no bound state for this potential. In Sec.~\ref{Sec.conclusion}, we will give some conclusions for this paper.  
\section{Exact solution}
\label{Sec1}
In this paper, we will consider the $\mathcal{PT}$ symmetric potential $V(x)=A[\sech(\lambda x)+i \tanh(\lambda x)]$, then the one-dimension Schr\"{o}dinger equation write as
\begin{align}
[\frac{d^2}{d x^2} +E-V(x)]\psi(x)=0\,.
\end{align}
Define $z=ie^{\lambda x}$, which transform the variable $x\in(-\infty,\infty)$ to a new variable $z\in(0,\infty)$,  
and introduce a function $u(z)$, define as $\psi(z)=(-z)^p u(z)$, by means of this substitution, we can rewrite the Schr\"{o}dinger equation as
\begin{align}
&z(1-z) u''(z)+[1 + 2p - (1 + 2p) z]u'(z)+\frac{ E-iA+p^2\lambda^2}{\lambda^2}u(z)=0\,.
\end{align} 
If and only if $E+iA+p^2\lambda^2=0$, Namely
\begin{align}
p=\pm\sqrt{-iA-E} /\lambda\,.
\end{align}
We can write the solution by comparing with hypergeometric equation
\begin{align}
z(1-z)u''(z)+[\gamma-(\alpha+\beta+1)z]u'(z)-\alpha\beta u(z)=0\,.
\end{align}
After simple calculation, we have
 \begin{align}
\alpha=p\pm\sqrt{iA-E} /\lambda\equiv p\pm q\,, \qquad \beta=2p-\alpha\,, \qquad \gamma=2p+1\,.
\end{align}  
Because of the commutative symmetry of $\alpha$ and $\beta$, without losing generality, we can choose $\alpha= p - q$.
From the hypergeometric functions theory, the wave function is given by
\begin{align}
\psi(z)&=(-z)^p[c_1F(\alpha,\beta,\gamma;z)+c_2(-z)^{1-\gamma}F(\alpha-\gamma+1,\beta-\gamma+1,2-\gamma;z)]\,.
\label{eq:6}  
\end{align}
Where $c_1,c_2$ are two arbitrary combination coefficients.
Due to the positive and negative sign of $p$ cause the exchange of left and right travelling waves of wave function.
In this we can choose $p=\sqrt{-iA-E} /\lambda$, we assume$\lambda>0$ for simplify.
\section{Scattering states}
\label{Sec2}
Considering the parametrization
\begin{align}
 \sqrt{E+iA}=k_R+ik_I=k\,.\,\qquad \sqrt{E-iA}=k'_R+ik'_I=k'
\end{align}
and assuming that $E,A$ are real numbers, the relations
\begin{align}
E=k_R^2-k_I^2={k'}_R^2-{k'}_I^2\,\qquad A=2k_Rk_I=-2{k'}_R{k'}_I\,. 
\end{align}
So we have $k^2=[(k')^2]^*$, which implies that $k'=\pm k^*$.

For $A>0$, choice $k'=k^*$ this mean $k_R=k'_R>0$ and $k_I=-k'_I>0$ (For $A<0$ the roles of $k$ and $k'$ are exchanged).
Let us analyse the asymptotic behaviour of solutions of (\ref{eq:6})  for $x\rightarrow -\infty$ (or $z\rightarrow 0$), we have 
\begin{align}
&\psi(x)_{-\infty}\sim c_1(-z)^p +c_2(-z)^{-p}=(-i)^pc_1e^{ikx} +(-i)^{-p}c_2e^{-ikx}\,.
\end{align}
So we have 
\begin{align}
a_{1-}=(-i)^{p}\,\qquad a_{2-}=0\,\qquad b_{1-}=0\,\qquad b_{2-}=(-i)^{-p},\,.
 \label{eq:a} 
\end{align}
where the notation can be found in \cite{Cannata:2006htc} or see appendix \ref{Sec.appendix1}.

And the asymptotic behaviour of the wave function $\psi(x)$ for $x\rightarrow \infty$ (or $z\rightarrow \infty$) is given 
\begin{align}
\psi(x)_{\infty} 
&=[c_1\frac{\Gamma(1+2p)\Gamma(-2q)}{\Gamma(p-q)\Gamma(1+p-q)}+c_2\frac{\Gamma(1-2p)\Gamma(-2q)}{\Gamma(-p-q)\Gamma(1-p-q)}](-z)^{-q}\nn\\
&+[c_1\frac{\Gamma(1+2p)\Gamma(2q)}{\Gamma(p+q)\Gamma(1+p+q)}+c_2\frac{\Gamma(1-2p)\Gamma(2q)}{\Gamma(-p+q)\Gamma(1-p+q)}](-z)^{q}\,.
\label{eq:c} 
\end{align}
where we used the formula 
\begin{align}
F(\alpha,\beta,\gamma;z)=\frac{\Gamma(\gamma)\Gamma(\beta-\alpha)}{\Gamma(\beta)\Gamma(\gamma-\alpha)}(-z)^{-\alpha}F(\alpha,\alpha-\gamma+1,\alpha-\beta+1;1/z)\nn\\
+\frac{\Gamma(\gamma)\Gamma(\alpha-\beta)}{\Gamma(\alpha)\Gamma(\gamma-\beta)}(-z)^{-\beta}F(\beta,\beta-\gamma+1,\beta-\alpha+1;1/z)\,.
\end{align}
We have similar notation, which are expressed 
\begin{align}
b_{1+}
&=(-i)^{-q}\frac{\Gamma(1+2p)\Gamma(-2q)}{\Gamma(p-q)\Gamma(1+p-q)}\,,\qquad
b_{2+}=(-i)^{-q}\frac{\Gamma(1-2p)\Gamma(-2q)}{\Gamma(-p-q)\Gamma(1-p-q)}\nn\\
a_{1+}&=(-i)^{-q}\frac{\Gamma(1+2p)\Gamma(2q)}{\Gamma(p+q)\Gamma(1+p+q)}\,,\qquad
a_{2+}=(-i)^{-q}\frac{\Gamma(1-2p)\Gamma(2q)}{\Gamma(-p+q)\Gamma(1-p+q)}\,.
 \label{eq:b} 
\end{align}
From Eq.~(\ref{eq:a}) and Eq.~(\ref{eq:b}), we have the transmission and reflection coefficients for a wave coming from the right are 
\begin{align}
&R_{R\rightarrow L}=(-i)^{2q}\frac{\Gamma(-p-q)\Gamma(1-p-q)\Gamma(2q)}{\Gamma(-2q)\Gamma(-p+q)\Gamma(1-p+q)}\,,\nn\\
&T_{R\rightarrow L}=(-i)^{q-p}\frac{\Gamma(-p-q)\Gamma(1-p-q)}{\Gamma(1-2p)\Gamma(-2q)}\,.
\end{align}
The transmission and reflection coefficients for a wave coming from the left are also given
\begin{align}
&R_{L\rightarrow R}=-
e^{ip\pi}\frac{\Gamma(1+2p)\Gamma(-p-q)\Gamma(1-p-q)}{\Gamma(p-q)\Gamma(1+p-q)\Gamma(1-2p)}\,,\nn\\
&T_{L\rightarrow R}=\fft{p}{q}T_{R\rightarrow L}\,.
\end{align}
we see that $R_{R\rightarrow L}\neq R_{L\rightarrow R}$, this reflects the handedness of the problem. As the imaginary part of potential asymptotically non-vanishing, that have 
$T_{L\rightarrow R}=\fft{p}{q}T_{R\rightarrow L}$.

Here we point out for $A>0$, if let $k'=-k^*$ this mean $k_R=-k'_R>0$ and $k_I=k'_I>0$ (For $A<0$ the roles of $k$ and $k'$ are exchanged).
The corresponding transmission and reflection coefficients is expressed as 
\begin{align}
&R_{R\rightarrow L}=e^{i\pi q}\frac{\Gamma(-p+q)\Gamma(1-p+q)\Gamma(-2q)}{\Gamma(2q)\Gamma(-p-q)\Gamma(1-p-q)}\,,\nn\\
&T_{R\rightarrow L}=e^{i\pi(p+q)/2}\frac{\Gamma(-p+q)\Gamma(1-p+q)}{\Gamma(1-2p)\Gamma(2q)}\,,\nn\\
&R_{L\rightarrow R}=-
e^{ip\pi}\frac{\Gamma(1+2p)\Gamma(-p+q)\Gamma(1-p+q)}{\Gamma(p+q)\Gamma(1+p+q)\Gamma(1-2p)}\,,\nn\\
&T_{L\rightarrow R}=-\fft{p}{q}T_{R\rightarrow L}\,.
\label{eq:rl} 
\end{align}
We can see that we just need to change $q$ to $-q$ in Eq.~(\ref{eq:rl}), the expression will return to the case of $k'=-k^*$. 
If we exchange p and q in transmission and reflection coefficients,
which will correspond to the situation when $A<0$.
\section{Bound states}
\label{Sec3}
Let us try to find the solutions of the eigenvalue problem  associated to this potential, One way to obtain those solutions is to analyze the behavior of the eigenfunctions in the limits $x\rightarrow -\infty$. For $A>0$, without loss of generality, we choose $\text{Re} p<0$, this corresponds to the case of $k_I=-k'_I>0$,
it is required that eigenfunctions square integrable, so we have $\psi(x)\sim z^{-p} F(\alpha-\gamma+1,\beta-\gamma+1,2-\gamma;z)$. Similarly, when $x\rightarrow \infty$,  see Eq.~(\ref{eq:c}), if $\text{Re} q>0$ we obtain bound states only when $-p+ q=-n,n=1,2,3,\cdots$, but in this case we have $-p+ q>0$, mean $n<0$. If $\text{Re} q<0$ we obtain bound states only when $-p- q=-n,n=1,2,3,\cdots$, but in this case we have $-p- q>0$, mean $n<0$, both of these are inconsistent with $n$ being a positive integer.  If we choose $\text{Re} p>0$, this corresponds to the case of $k_I=-k'_I<0$, we will have  bound states conditions $p+ q=-n,n=1,2,3,\cdots$ for $\text{Re} q>0$, or $p- q=-n,n=1,2,3,\cdots$ for $\text{Re} q<0$,we also found these cases are inconsistent with $n$ being a positive integer.

By the way, for $A<0$, corresponds to exchanging the momenta $k$ and $k'$, also corresponds to exchanging the $p$ and $q$.  Choose $\text{Re} q<0$, we have $\psi(x)\sim z^{-q} F(\alpha-\gamma+1,\beta-\gamma+1,2-\gamma;z)$ with $x\rightarrow -\infty$. When $x\rightarrow \infty$, if $\text{Re} p>0$ we obtain bound states only when $-q+ p=-n,n=1,2,3,\cdots$, but in this case $-q+ p>0$, mean $n<0$. If $\text{Re} p<0$ we have bound states only when $-p- q=-n,n=1,2,3,\cdots$, but in this case $-p- q>0$, mean $n<0$, both of these are inconsistent with $n$ being a positive integer.  If we choose $\text{Re} q>0$, this corresponds to the case of $k_I=-k'_I<0$, we will have  bound states conditions $p+ q=-n,n=1,2,3,\cdots$ for $\text{Re} p>0$, or $q- p=-n,n=1,2,3,\cdots$ for $\text{Re} p<0$,we also found these cases are inconsistent with $n$ being a positive integer. So we have conclusion that there is no bound state solution for potential $V(x)=A[\sech(\lambda x)+i \tanh(\lambda x)]$.

\section{Conclusion}
\label{Sec.conclusion}
In this work, we consider to solve Schr\"{o}dinger equations with the potential $V(x)=A[\sech(\lambda x)+i \tanh(\lambda x)]$,
and calculated the transmission and reflection coefficients of this potential. 
For this $\mathcal{PT}$ symmetric potential, the wave numbers are complex and the asymptotic wave numbers  in $x\rightarrow \infty$ is different in $x\rightarrow -\infty$, cause the transmission coefficient changes a complex phase factor.
The handedness effect was found by changing the direction of the incoming wave, where the reflection coefficient changes.
By the way, we also can consider the $\mathcal{PT}$ symmetric potential $V(x)=A[\sech(\lambda x)-i \tanh(\lambda x)]$,
that only need redefine $z=-ie^{\lambda x}$, and  change the $iA$ to $-iA$ in the solutions of (\ref{eq:6}).
Obviously, there is no bound state for this potential.

\section*{Acknowledgement}
This work was supported by the Guangxi Scientific Programm Foundation under grant No. 2020AC20014, the Scientific Research Foundation of Guilin University of Technology under grant No. GUTQDJJ2019206.

\appendix
\section{Transmission and reflection coefficients}
\label{Sec.appendix1}
Consider the time independent Schr\"{o}dinger  equation satisfied by 
\begin{align}
H\psi(x)=[-\frac{d^2}{d x^2}+V(x)]\psi(x)=k^2\psi(x)\,.
\end{align}
With the asymptotic states (finite-range local potential) can be expressed at $x\rightarrow \pm\infty$ as
\begin{align*}
\begin{split}
\psi(x)= \left \{ 
\begin{array}{ll} 
    A_-e^{ikx}+B_-e^{-ikx},   & x\rightarrow -\infty\\ 
    A_+e^{ikx}+B_+e^{-ikx},  & x\rightarrow -\infty
\end{array} 
\right. 
\end{split} 
\end{align*}
Schr\"{o}dinger  equation admits a general solution written as two linear independent solutions, $F_1 (x)$ and $F_2 (x)$,
with non-zero Wronskian, whose asymptotic expressions as:
\begin{align}
&\lim_{x\rightarrow \pm\infty}F_1=a_{1\pm}e^{ikx}+b_{1\pm}e^{-ikx}\,,\nn\\
&\lim_{x\rightarrow \pm\infty}F_2=a_{2\pm}e^{ikx}+b_{2\pm}e^{-ikx}\,.
\end{align}
which related to the asymptotic amplitudes $A_{\pm}$ and $B_{\pm}$ 
\begin{align}
&A_{\pm}=\alpha a_{1\pm}+\beta a_{2\pm}\,,\nn\\
&B_{\pm}=\alpha b_{1\pm}+\beta b_{2\pm}\,.
\end{align}
when the wave moving from left to right, the transmission and reflection coefficients are written as
\begin{align}
&T_{L\rightarrow R}=\fft{a_{2+}b_{1+}-a_{1+}b_{2+}}{a_{2-}b_{1+}-a_{1-}b_{2+}}\,,\nn\\
&R_{L\rightarrow R}=\fft{b_{1+}b_{2-}-b_{1-}b_{2+}}{a_{2-}b_{1+}-a_{1-}b_{2+}}\,.
\end{align}
when the wave moving from right to left, the transmission and reflection coefficients are written as
\begin{align}
&T_{R\rightarrow L}=\fft{a_{2-}b_{1-}-a_{1-}b_{2-}}{a_{2-}b_{1+}-a_{1-}b_{2+}}\,,\nn\\
&R_{L\rightarrow R}=\fft{a_{1+}a_{2-}-a_{1-}a_{2+}}{a_{2-}b_{1+}-a_{1-}b_{2+}}\,.
\end{align}

\end{CJK}

\end{document}